%% file: main.tex
\documentclass[runningheads]{llncs}

\usepackage{listings}
\usepackage{booktabs}
\usepackage{comment}
\usepackage[dvipsnames]{xcolor}
\definecolor{codegreen}{rgb}{0,0.6,0}
\definecolor{codegray}{rgb}{0.5,0.5,0.5}
\definecolor{codepurple}{rgb}{0.58,0,0.82}
\definecolor{backcolour}{rgb}{0.95,0.95,0.92}
\lstdefinestyle{myStyle}{
    belowcaptionskip=1\baselineskip,
    breaklines=true,
    frame=none,
    numbers=left,
    basicstyle=\footnotesize\ttfamily,
    keywordstyle=\bfseries\color{green!40!black},
    commentstyle=\itshape\color{purple!40!black},
    identifierstyle=\color{blue},
}
\usepackage{multirow}
\usepackage{tabularx}
\usepackage{caption}
\usepackage{subcaption}
\usepackage{graphicx}
\usepackage{amsmath}
\usepackage{lineno}
\usepackage{ulem}

\nolinenumbers

\AtBeginDocument{%
  \providecommand\BibTeX{{%
    \normalfont B\kern-0.5em{\scshape i\kern-0.25em b}\kern-0.8em\TeX}}}

\newcommand{\simtask}[0]{\texttt{Simulation}}
\newcommand{\aggtask}[0]{\texttt{Aggregation}}
\newcommand{\mltask}[0]{\texttt{Training}}
\newcommand{\inftask}[0]{\texttt{Inference}}
\newcommand{\ie}[0]{\textit{i.e.}}
\newcommand{\eg}[0]{\textit{e.g.}}
\newcommand{\etc}[0]{\textit{etc.}}

\newcommand{\taggr}[0]{t_{\text{Aggr}}}
\newcommand{\tml}[0]{t_{\text{Train}}}
\newcommand{\tinf}[0]{t_{\text{Infer}}}
\newcommand{\tseq}[0]{t^{\text{Seq}}}
\newcommand{\tasync}[0]{t^{\text{Async}}}
\newcommand{\doadep}[0]{\text{DOA}_\text{dep}}
\newcommand{\doares}[0]{\text{DOA}_\text{res}}

\newcommand{\NOTE}[1]{\phantom{}\begingroup\relax\ifmmode\boldmath\else\bfseries\fi\color{Cerulean}\ignorespaces#1\ignorespaces\endgroup}
\newcommand{\TODO}[1]{\phantom{}\begingroup\relax\ifmmode\else\sffamily\fi\color{BurntOrange}\ignorespaces#1\ignorespaces\endgroup}
\newcommand{\FIXME}[1]{\phantom{}\begingroup\relax\ifmmode\boldmath\else\bfseries\sffamily\fi\color{Red}\ignorespaces#1\ignorespaces\endgroup}
\newcommand{\FIXED}[1]{\phantom{}\begingroup\relax\ifmmode\else\sffamily\fi\color{Green}\ignorespaces#1\ignorespaces\endgroup}
\newcommand{\DELETE}[1]{\phantom{}\begingroup\relax\ifmmode\else\sffamily\fi\color{Red}\ifmmode\text{\sout{\ensuremath{#1}}}\else\sout{\ignorespaces#1\ignorespaces}\fi\endgroup}

\newcommand{\UP}{\vspace*{-1.0em}}

\newif\ifdraft{}
\drafttrue{}

\ifdraft{}
  \newcommand{\jhanote}[1]{ {\textcolor{red} { ***shantenu: #1 }}}
  \newcommand{\mtnote}[1]{ {\textcolor{orange} { ***matteo: #1 }}}
  \newcommand{\vrpnote}[1]{ {\textcolor{blue} { ***[vrp]: #1 }}}
  \newcommand{\ozgurnote}[1]{ {\textcolor{green} { ***[ozgur]: #1 }}}
\else
  \newcommand{\jhanote}[1]{}
  \newcommand{\mtnote}[1]{}
  \newcommand{\vrpnote}[1]{}
  \newcommand{\ozgurnote}[1]{}
\fi

\begin{document}


\title{Asynchronous Execution of Heterogeneous Tasks in ML-driven HPC Workflows}

\author{
    Vincent R. Pascuzzi\inst{1}\orcidID{0000-0003-3167-8773} \and
    Ozgur O. Kilic\inst{1}\orcidID{0000-0003-2129-408X} \and
    Matteo Turilli\inst{1,2}\orcidID{0000-0003-0527-1435} \and
    Shantenu Jha\inst{1,2}\orcidID{0000-0002-5040-026X}
}
\institute{
    Brookhaven National Laboratory, Upton, NY 11973, USA
    \email{vrpascuzzi@gmail.com, \{okilic, mturilli\}@bnl.gov}
    \and
    Rutgers, the State University of New Jersey, Piscataway, NJ 08854, USA
    \email{shantenu.jha@rutgers.edu}
}

\maketitle

\begin{abstract}
Heterogeneous scientific workflows consist of numerous types of tasks that
require executing on heterogeneous resources. Asynchronous execution of
those tasks is crucial to improve resource utilization, task 
throughput and reduce workflows' makespan. Therefore,
middleware capable of scheduling and executing different task types across
heterogeneous resources must enable asynchronous execution of tasks. In this
paper, we investigate the requirements and properties of the asynchronous task
execution of machine learning (ML)-driven high performance computing (HPC) 
workflows. We model the degree of asynchronicity
permitted for arbitrary workflows and propose key metrics that can be used to
determine qualitative benefits when employing asynchronous execution. Our
experiments represent relevant scientific drivers, we perform them at scale on
Summit, and we show that the performance enhancements due to asynchronous 
execution are consistent with our model.

\end{abstract}

\thispagestyle{plain}
\pagestyle{plain}

\section{Introduction}\label{sec:intro}
\input{intro-2}

\section{Related Work}\label{sec:related}
\input{related}

\section{Motivation}\label{sec:motivation}
\input{motivation}

\section{Design and Implementation}\label{sec:design}
\input{design}

\section{Workload-Level Asynchronicity}\label{sec:conds}
\input{conditions_and_benefits}

\section{Experiments}\label{sec:experiments}
\input{experiments}

\section{Performance Characterization}\label{sec:perf}
\input{performance}

\section{Conclusions}\label{sec:conclusions}
\input{conclusions}

\section*{Acknowledgements}
\footnotesize{This work was supported by the ECP CANDLE and ExaWorks projects,
the DOE HEP Center for Computational Excellence at BNL under B\&R KA2401045, as
well as NSF-1931512 (RADICAL-Cybertools). We thank Andre Merzky, Li Tan, and
Ning Bao for useful discussions and support. We acknowledge Arvind Ramanathan,
Austin Clyde, Rick Stevens and  Ian Foster for useful discussions and
co-development of DeepDriveMD. This research used resources at OLCF ORNL, which
is supported by the Office of Science of the U.S. Department of Energy under
Contract No. DE-AC05-00OR22725.}

\bibliographystyle{splncs04}
\bibliography{radical,main}


\end{document}

%% file: intro-2.tex

Scientific discovery increasingly requires sophisticated and scalable workflows.
Introducing machine learning (ML) models into traditional high performance
computing (HPC) workflows has been an enabler of highly accurate modeling and a
promising approach for significant scientific
advances~\cite{casalino2020aidriven}. Those workflows often comprise hundreds to
thousands of heterogeneous tasks~\cite{covidisairborne2021ijhpca}
with diverse dimensions of heterogeneity: \textbf{implementation} (\eg,
executables or functions/methods in an arbitrary language), \textbf{resource
requirements} (\eg, executing on central processing units (CPUs)  
and/or graphics processing units (GPUs),
computational or
data-intensive operations), \textbf{duration} (\eg, from less than a second to
many hours), and \textbf{size}, (\eg, from a core to many cores over one or more
compute nodes).
Specifically, for ML-driven
HPC workflows, task heterogeneity increases
with
GPU and multi-node message passing interface (MPI)~\cite{gropp1999using} tasks alongside traditional CPU tasks. ML tasks usually
require the execution of high-throughput function calls, often implemented in an
interpreted language such as Python.

Typically, in ML-driven HPC workflows, simulation tasks generate data, and ML
tasks use that data for training or inference, while inference results are used
to guide the next set of simulations~\cite{ward2021colmena,brace2021achieving},
creating a dependence between the two types of tasks. A naive approach in which
all the simulations run before the ML tasks, or all simulations pause while inference
(or training) is ongoing, leads to poor resource utilization, \eg, while the ML
tasks run,
a large amount of the allocated HPC resource idles.
Further, continuous learning, surrogate re-training, and optimal execution of
campaigns are becoming increasingly important and prevalent, making
asynchronous execution of simulations and ML phases of the workflow unavoidable,
and
bulk-synchronous parallel
(BSP) executions with their hard temporal separation of simulations and ML phases untenable.

There are multiple levels at which asynchronous execution is possible
and should be supported: workflow-level, \ie, executing independent workflows
asynchronously while preserving dependencies within each workflow. For
example, in Ref.~\cite{saadi2021impeccable}, different workflows can be
executed without waiting for all instances of one workflow to finish;
workload-level, \ie, executing a set of heterogeneous tasks whose dependencies
have been resolved asynchronously, where the set of tasks might be derived
from multiple workflows; and, task-level, \ie, executing each independent task
asynchronously. 
As a consequence, the middleware that manages the execution of the workflow
application has to be able to
asynchronously
schedule, place, and execute simulations and ML tasks. For example, MPI executables
alongside Python functions with varying resource
requirements and execution lifetimes, the former
for hours, the latter for less than a second. For the task-execution middleware to do so, in a way that
maximizes resource utilization while minimizing the workflow makespan, requires
it to explicitly support the asynchronous execution of heterogeneous tasks.
Resource utilization and efficiency depend upon the specifics of each workflow,
workload mix (type of tasks and their individual properties), and how
effectively the task-execution middleware manages the heterogeneity of the tasks
and resources.



In this paper, we focus on the resource management challenges of
executing tasks with diverse resource requirements. We discuss conditions for
asynchronous execution and offer a quantitative basis and experimental
characterization of the improvements it can bring in terms of resource
utilization and task throughput, and thus makespan. We focus on the asynchronous
execution of ML-driven heterogeneous HPC workflows, offering four main
contributions: (1) an asynchronous implementation of
DeepDriveMD~\cite{brace2021achieving,lee2019deepdrivemd}---a framework to
execute ML-driven molecular-dynamics workflows on HPC platforms at scale; (2)
a performance evaluation of asynchronous DeepDriveMD; (3) a model of
asynchronous behavior; and (4) a general performance evaluation of that model
for workflows with a varying degree of asynchronous execution.

In \S\ref{sec:related}, we discuss several ML-driven HPC workflows and
workflow systems, as generalized motivation and the importance of asynchronous
execution. In \S\ref{sec:design}, we define asynchronous execution and its
relationship with data and control flow when designing a workflow application.
In \S\ref{sec:conds}, we present conditions
and the motivation
for workflows-level asynchronicity with the concept of task execution time
masking, which together can be used to estimate the expected relative
improvement in adopting asynchronous execution pattern with respect to
sequential execution for arbitrary workflows. In \S\ref{sec:experiments},
we formalize the notion of asynchronous execution by modeling its behavior
and performance for task throughput and resource utilization metrics. In
\S\ref{sec:perf}, we compare synchronous and asynchronous execution of
DeepDriveMD, measuring the latter's benefits. We then generalize those
results, offering an experimental characterization of the performance gains
provided by different degrees of heterogeneity for a given workflow. Finally, in
\S\ref{sec:conclusions}, we discuss the implications of our model and performance
analysis for the design of workflow applications with heterogeneous tasks when
executed on HPC resources.

%% file: related.tex

Asynchronicity spans multiple levels of an HPC stack and affects many areas of a
distributed execution. Asynchronous execution, communication, and coordination
are implemented and leveraged in many runtime systems, both to distribute
workload among executors and/or coordinate the messaging among executing
processes. Uintah~\cite{meng2012uintah}, Charm++~\cite{kale1993charm++},
HPX~\cite{kaiser2014hpx},
Legion~\cite{bauer2012legion},
and PaRSEC~\cite{bosilca2013parsec} are all examples of runtime using
asynchronicity to coordinate execution.

Contrary to the runtime system listed above, in this paper, we focus on
asynchronous execution for ML-driven HPC workflows which consist of tasks with
diverse dimensions of heterogeneity, \eg, implementation, resource requirements,
duration, or size. We assume tasks to be black boxes. Therefore, in our
modeling, we abstract away the details of asynchronous communication as data
dependencies are assumed to be satisfied either before or during execution.


%



PATHWAYS~\cite{barham2022pathways} focuses on the asynchronous distribution of
dataflow to enable parallel execution of ML tasks via a gang-scheduling
approach. PATHWAYS overlaps computation with data coordination by dispatching
data asynchronously, showing how that increases resource utilization while
reducing makespan. Our work differs from PATHWAYS as we focus on asynchronous
execution of heterogeneous tasks instead of asynchronous data distribution.


DeepHyper~\cite{balaprakash2018deephyper} shows how asynchronous execution can
benefit the hyper-parameter search for deep neural networks (DNN). They implement
task-level asynchronicity by evaluating partial results
instead of waiting for the complete results to be ready, showing how that helps
to increase
resource utilization. While specialized task execution models could benefit
hyper-parameter search in DNN, many dimensions of heterogeneity need to be
considered to improve the performance of ML-driven heterogeneous HPC workflows.
For this reason, we are focusing on a generalized asynchronous modeling and
execution of heterogeneous HPC workflows.

Significant examples of ML-driven HPC workflow solutions include
Colmena~\cite{ward2021colmena},
Mummi/Merlin~\cite{peterson2022enabling,di2019massively,bhatia2021generalizable},
and Proxima~\cite{zamora2021proxima}. Those solutions and the applications they
support involve running a time-varying heterogeneous mix of HPC and ML tasks. As
such, they could benefit from programmatic asynchronous execution, but they lack
specific constructs and tools to assess the performance improvement that an
asynchronous implementation would offer.

Further examples of ML-driven HPC workflow can be found
in Refs.~\cite{saadi2020impeccable,bhati2021pandemic}, both of which use
DeepDriveMD~\cite{lee2019deepdrivemd,brace2021achieving} to couple ML and
molecular dynamic simulations. The performance characterization of DeepDriveMD
established that, when feasible, asynchronous execution of ML and simulation
tasks can improve both resource utilization and task throughput. Those
improvements are possible due to the heterogeneity of ML and simulation tasks in
all dimensions mentioned in \S\ref{sec:intro}. Nonetheless, the implementation
of asynchronous execution depends on the requirements of the task utilized and
cannot be applied to all task types executed via DeepDriveMD.


%% file: motivation.tex
Increasing heterogeneity in scientific workflows makes asynchronous
execution of heterogeneous workloads unavoidable. We can use the different
ML-Driven workflow implementations for neutron scattering data analysis
presented in~\cite{wang2023IPDPS} as a road map for achieving higher resource utilization and lower makespan with
increased degrees of asynchronicity. Their synchronous baseline, where the entire
simulation executes before ML execution starts, suffers from memory bottleneck
due to the amount of data produced. Serial workflow reduces memory pressure by
partitioning those tasks (simulation and ML) into so-called `phases'; however,
its synchronous execution of heterogeneous workloads could not get maximum
resource utilization. Asynchronous execution (`parallel workflow') improves
resource utilization while reducing the makespan. As mentioned in their future
work, the next step will be to use an adaptive execution to reduce their makespan further.
It is important to understand that we need an asynchronous execution model and
framework to run the increasingly heterogeneous scientific workflow efficiently. We
already mentioned existing systems that show the importance of different degrees
of asynchronicity and scientific workflows that can benefit from asynchronous
execution in \S\ref{sec:related}.


We investigate asynchronous execution of ML-driven HPC workflows, i.e.,
workflows with heterogeneous tasks. In this context, we define asynchronous
execution as one in which multiple heterogeneous tasks execute independently.
Those independent tasks can execute asynchronously either at the same time or
back-to-back, depending on resource availability. For this reason, we argue
there exist two conditions for workload-level asynchronicity (WLA): (1)
inter-task dependencies, and (2) resource availability.

Our main objective is to understand how to reduce the workflow makespan
and improve resource utilization,
utilizing asynchronous execution. We need asynchronicity
because of the characteristics of ML-driven HPC workflows: (1) their multiple
dimensions of heterogeneity (implementation, duration, size, and resource
requirements); and (2) their adaptive, dynamic, and changing workloads.
%
WLA alone does not guarantee better resource utilization and/or makespan as
the realized asynchronicity of the execution depends on task and dependency
graph properties, e.g., task execution time (TX) and the number of independent
branches in the graph. Thus, we explain the effect of the asynchronous execution
of heterogeneous workflow tasks on the makespan of the entire workflow and its
correlation with resource utilization.





%% file: design.tex


Computationally, HPC workflows often require the adoption of two task execution
paradigms: control-flow and data-flow. In the control-flow paradigm, executing a
workflow requires explicit control over distinct task instances. Each task
launches with an input data set, performs a fixed amount of computation on that
input, and exits execution. In the data-flow paradigm, workflow execution is
defined in terms of data dependencies among tasks, establishing what data each
task has to share with another task.

Design-wise, control-flow and data-flow paradigms are not mutually exclusive.
Workflow engines can independently manage data staging, inter-task
communication, task execution, and task processing. Task data requirements,
i.e., that a task can start processing only when specific input data are
available, can be equally satisfied by diverse designs: (1) Staging input data
before executing a task; or (2) Communicating (\eg, streaming or sending a
message with a data location) the input data to the task when that task is
already executing. Note that executing a task is different from having that task
starting to process input data: a task can be executing but also idle.

Workflows can benefit from the concurrent use of both control and data flows
when they require executing heterogeneous tasks. Some of those tasks may have to
communicate their input data to each other or may idle while waiting for input
data to be available. Other tasks may instead be launched only when data are
already available and terminated after those data have been processed. For
example, a certain type of tasks (\eg, MPI simulations) may occupy so many
resources that it would be too costly to keep them executing while idling.
Inversely, some other types of task (\eg, ML inference) may require limited and
specialized resources like GPUs, execute frequently and for only few seconds and
thus be too costly to launch and terminate every time they need to run.

We explore the design space of control and data flows by modeling asynchronous
execution of ML-driven HPC workflows. Workflow execution is asynchronous when
multiple \textbf{heterogeneous tasks execute independently} on the available
resources. A task executes independently when it does not need to coordinate
with other tasks to proceed with its execution.

Note that independence here pertains to tasks that are executing, not to the
dependency among tasks before their execution. For example, a task may depend on
data produced by another task but, when executing, those data are already
available and thus their availability does not require coordination among
executing tasks. In this context, it is crucial to distinguish between tasks
that are \textit{running} and those that are \textit{executing}. When running, a
task is not necessarily performing any useful calculation as it may be idling
(\ie, wasting resources) until triggered to do so. Instead, an executing task is
necessarily performing useful work that advances the workflow to its goal.

\begin{figure}[!ht]
     \centering
     \begin{subfigure}[b]{0.33\columnwidth}
         \centering
         \includegraphics[width=\columnwidth]{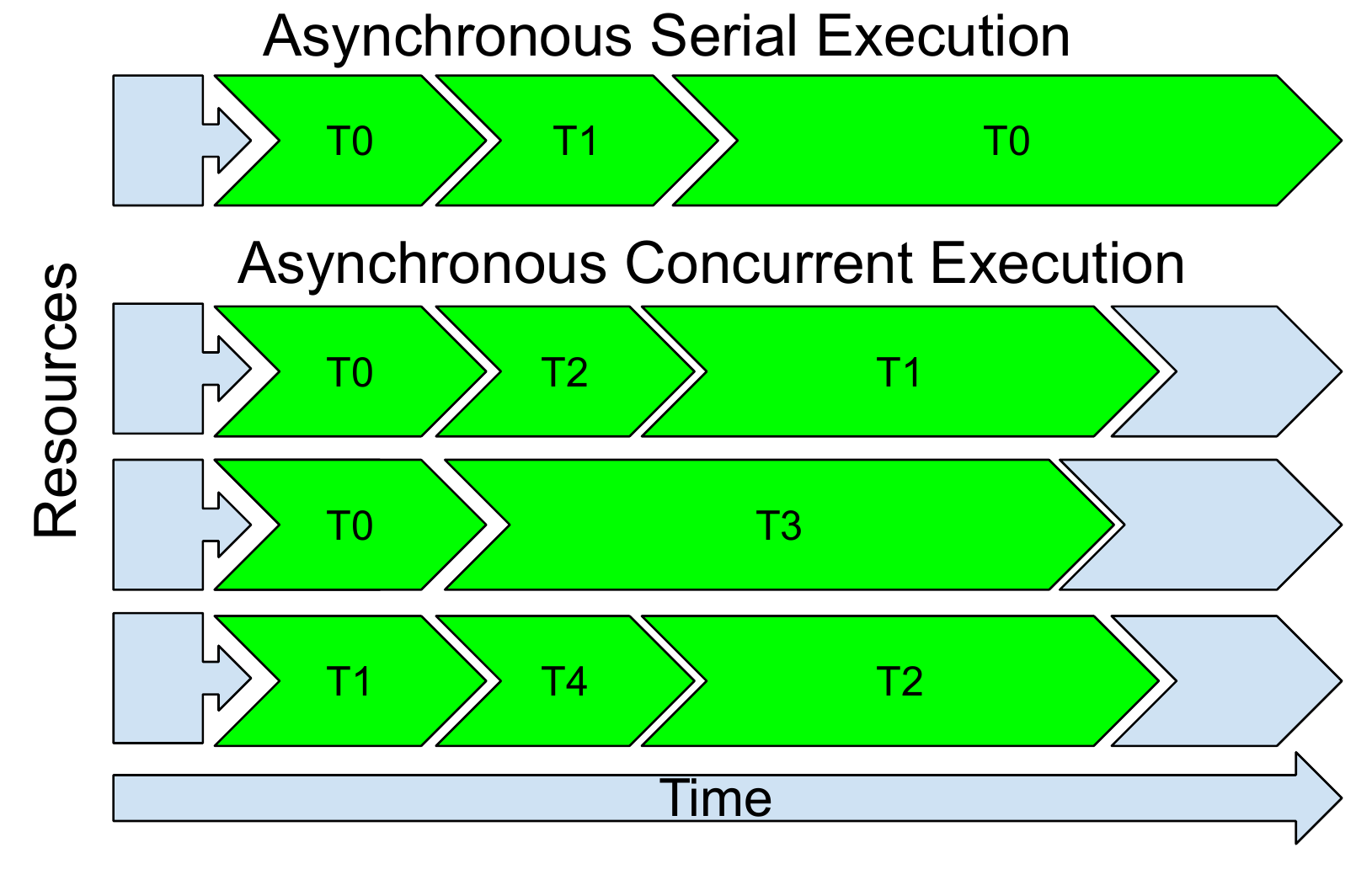}
         \caption{Async. Execution}
         \label{sfig:asyncexec}
     \end{subfigure} %
     \hfill
     \begin{subfigure}[b]{0.33\columnwidth}
         \centering
         \includegraphics[width=\columnwidth]{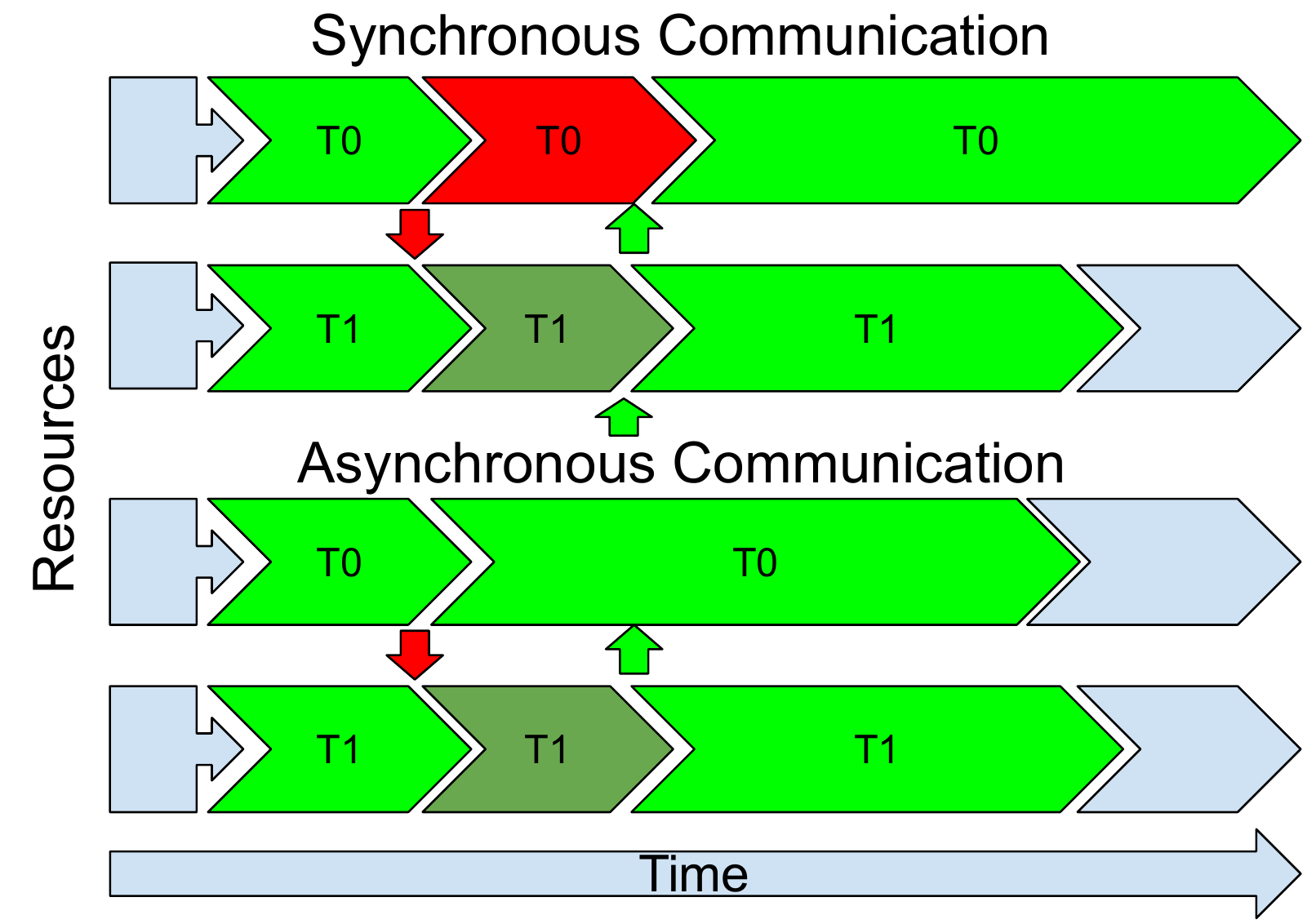}
         \caption{Async. Communication}
         \label{sfig:asynccomm}
     \end{subfigure}%
     \hfill
     \begin{subfigure}[b]{0.33\columnwidth}
         \centering
         \includegraphics[width=\columnwidth]{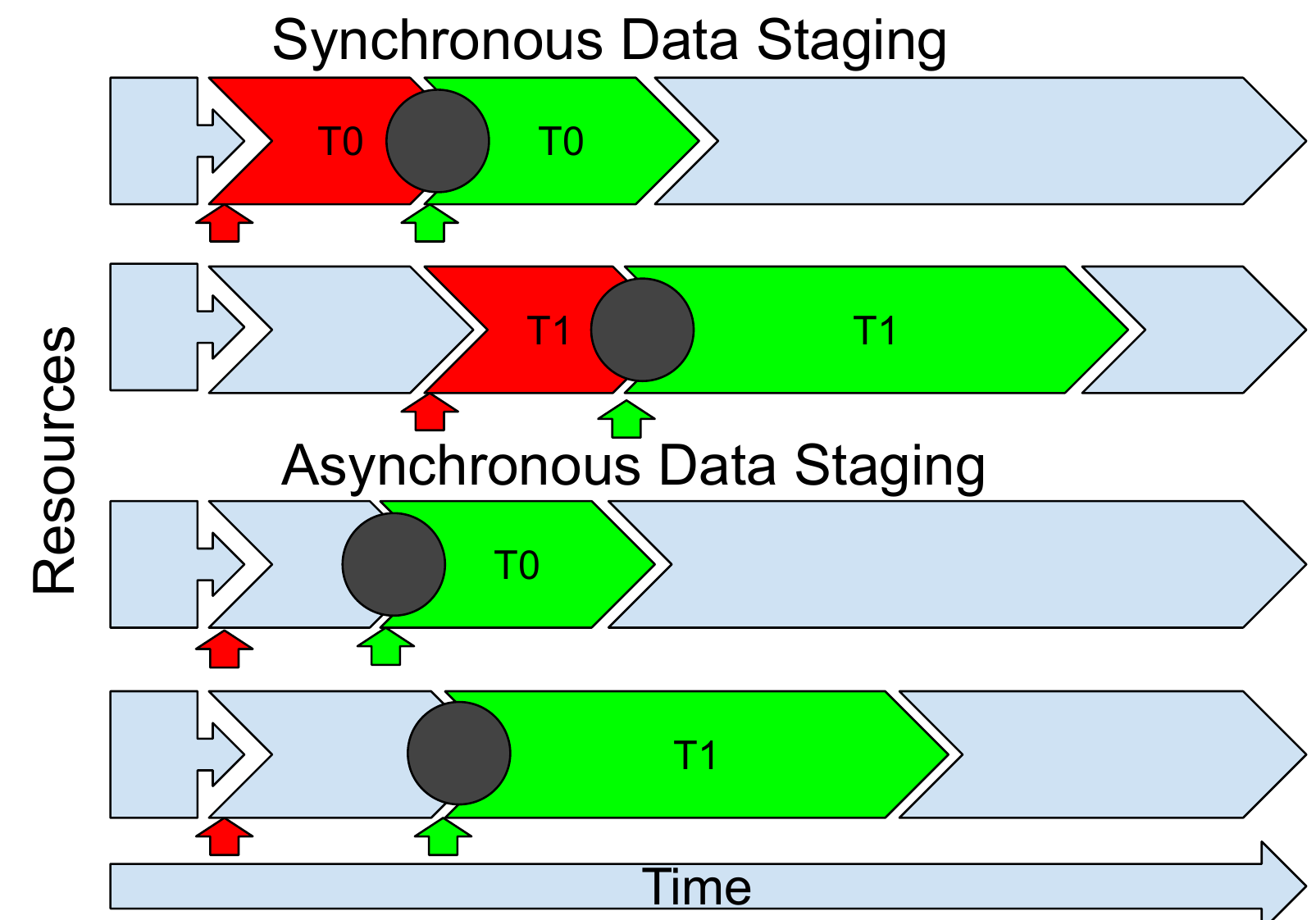}
         \caption{Async. Data Staging}
         \label{sfig:asyncdata}
     \end{subfigure}
        \caption{Types of Asynchronicity. Fig~\ref{sfig:asyncexec} shows two
        asynchronous executions: Top, T0 and T1 execute
        asynchronously and serially;
        bottom, T0 and T1 execute asynchronously and concurrently and  
        T0 and T2 execute synchronously. 
        Note that we used DG in Figure~\ref{sfig:graph-doa4}, assuming that 
        T0 and T1 do not have any dependencies and T2 depends on T0.
        Fig~\ref{sfig:asynccomm} shows asynchronous and synchronous
        communication: \textcolor{red}{red} arrows are sent messages and
        \textcolor{green}{green} arrows are received messages;
        \textcolor{red}{red}  bars
        indicate synchronous tasks that are waiting for their messages to be
        exchanged before resuming execution.
        and \textcolor{OliveGreen}{dark green} bar is the time taken to execute
        incoming request before replying.
        Fig~\ref{sfig:asyncdata} shows
        asynchronous and synchronous data staging: 
        \textcolor{red}{red} arrows show when data staging started and 
        \textcolor{green}{green} arrows show when data is ready;
        \textcolor{red}{red} bars indicate synchronous tasks
        that cannot be executed before data become available. 
        \label{fig:async-types}}
\end{figure}

Figure~\ref{fig:async-types} shows different types of asynchronicity.
Asynchronous execution~\ref{sfig:asyncexec} should not be confused with
`asynchronous data staging'~\ref{sfig:asyncdata} or the more common
`asynchronous communication'~\ref{sfig:asynccomm}. The former indicates that
each task's input data is staged independently; the latter that communications
among sender and receiver do not need to be coordinated. Independent tasks may
require staged files or asynchronous communication but, in this paper, we
abstract those details, focusing on modeling and characterizing asynchronous
execution as defined above.

Exemplars in
Refs.~\cite{saadi2020impeccable,bhati2021pandemic,brace2021achieving} require a
general-purpose approach to asynchronous task execution in DeepDriveMD. Such an
approach should not depend on the specific implementation of each task, remains
agnostic towards the communication and data capabilities of each task and makes
no assumptions about streaming capabilities. DeepDriveMD must account for tasks
that have to be executed sequentially, those that can concurrently execute and
communicate or exchange data, and those that have no relationship and can be
executed either concurrently or sequentially.

We extended DeepDriveMD~\cite{brace2021achieving} to implement asynchronous
executions of heterogeneous workflows. DeepDriveMD can now manage an arbitrary
number of task types, each requiring an arbitrary amount of CPU cores and/or
GPUs. Each task 
can be fully independent, dependent on the output
of another task or dependent on communicating with one or more other tasks.
The asynchronous execution of DeepDriveMD can be configured to execute tasks
concurrently or sequentially, considering their dependency constraints.
As with the previous version, DeepDriveMD is implemented as a thin layer on top
of the RADICAL-EnsembleToolkit (EnTK) workflow engine~\cite{entk2016bala}. EnTK
keeps track of task dependencies and submits them for execution to
RADICAL-Pilot~\cite{merzky2015radical}, a pilot system designed to execute
heterogeneous workloads on HPC platforms.

%% file: conditions_and_benefits.tex




In this section, we investigate the details of two conditions for workload-level
asynchronicity (WLA): (1) inter-task dependencies and (2) resource availability.
We explore  the benefits of the asynchronous execution of heterogeneous
workflows by focusing on makespan and resource utilization.

\subsection{Condition I: Inter-Task Dependencies}\label{ssec:cond-dependencies}

Asynchronicity of ML-enabled workflows depends on executing independent
heterogeneous tasks. We consider tasks to be black boxes, i.e., we do not make
assumptions about their inner working, including potential runtime inter-task
communication. Accordingly, we consider only data dependencies among tasks,
where data are consumed and produced as tasks' inputs and outputs.
Consistent with common practice, we represent task dependencies of a workflow as
a directed acyclic graph (DAG), where edges are data dependencies and nodes are
tasks.

\begin{figure}[!ht]
     \centering
     \begin{subfigure}[b]{0.25\columnwidth}
         \centering
         {\raisebox{30pt}{\includegraphics[width=\columnwidth]{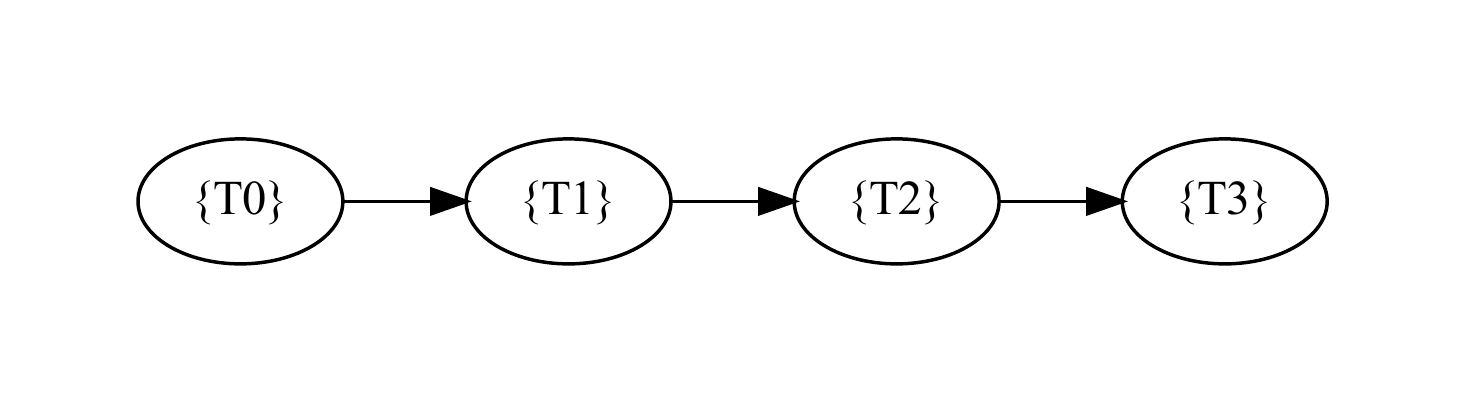}}}
         \caption{$\doadep = 0$}
         \label{sfig:graph-chain}
     \end{subfigure}%
     \hfill
     \begin{subfigure}[b]{0.19\columnwidth}
         \centering
         \includegraphics[width=\columnwidth]{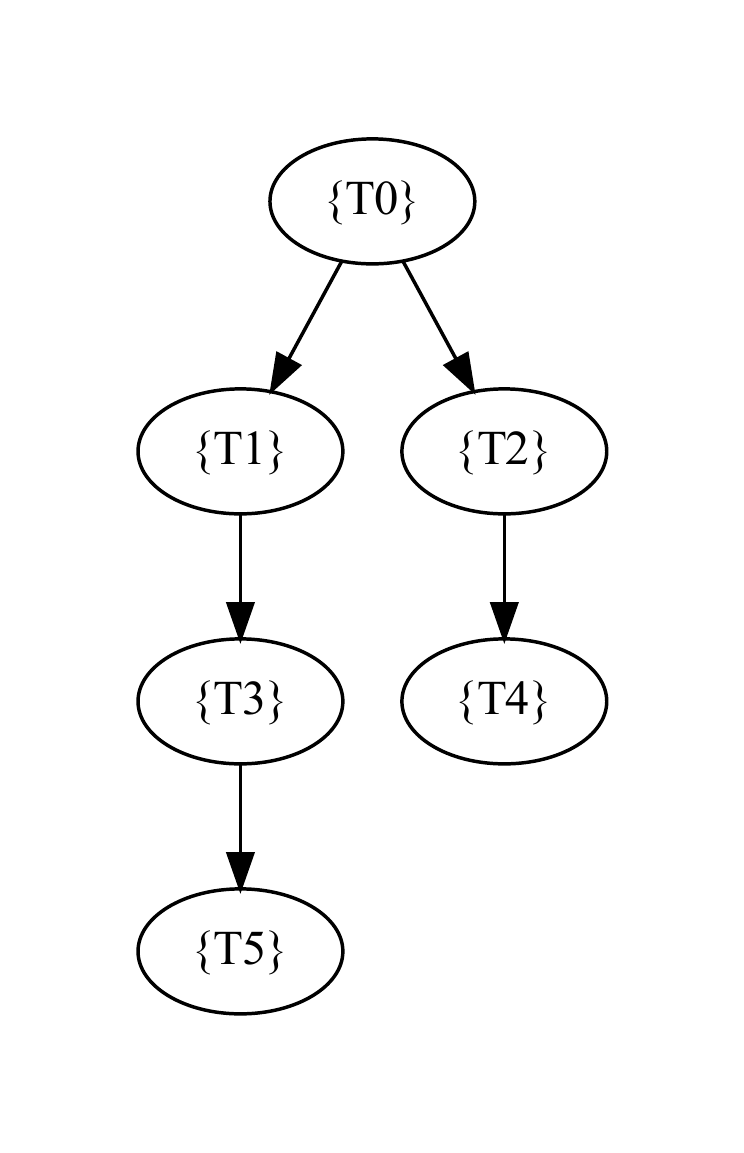}
         \caption{$\doadep = 1$}
         \label{sfig:graph-doa1}
     \end{subfigure} %
     \hfill
     \begin{subfigure}[b]{0.30\columnwidth}
         \centering
         {\raisebox{15pt}{\includegraphics[width=\columnwidth]{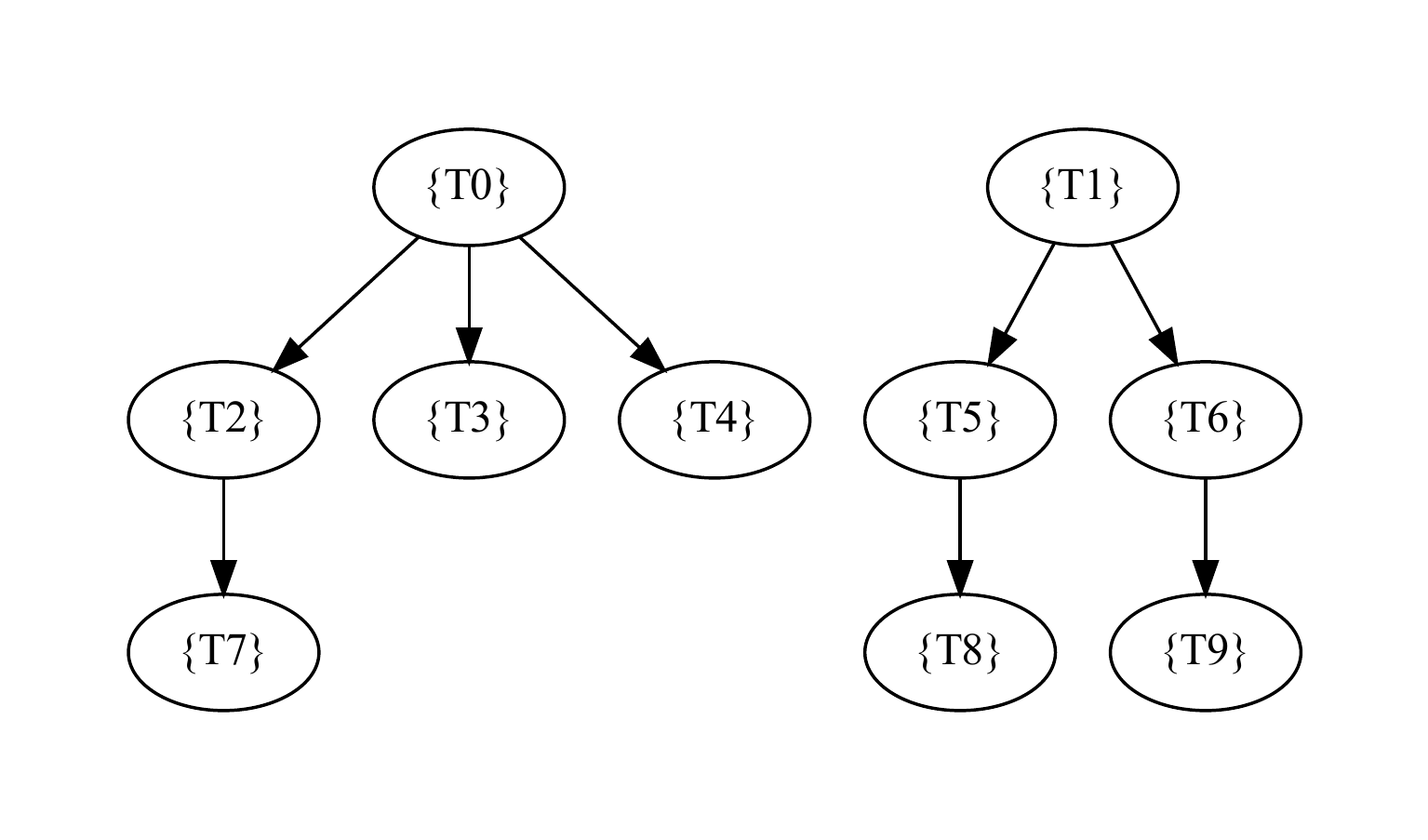}}}
         \caption{$\doadep = 4$}
         \label{sfig:graph-doa4}
     \end{subfigure}%
     \hfill
     \begin{subfigure}[b]{0.25\columnwidth}
         \centering
         {\raisebox{30pt}{\includegraphics[width=\columnwidth]{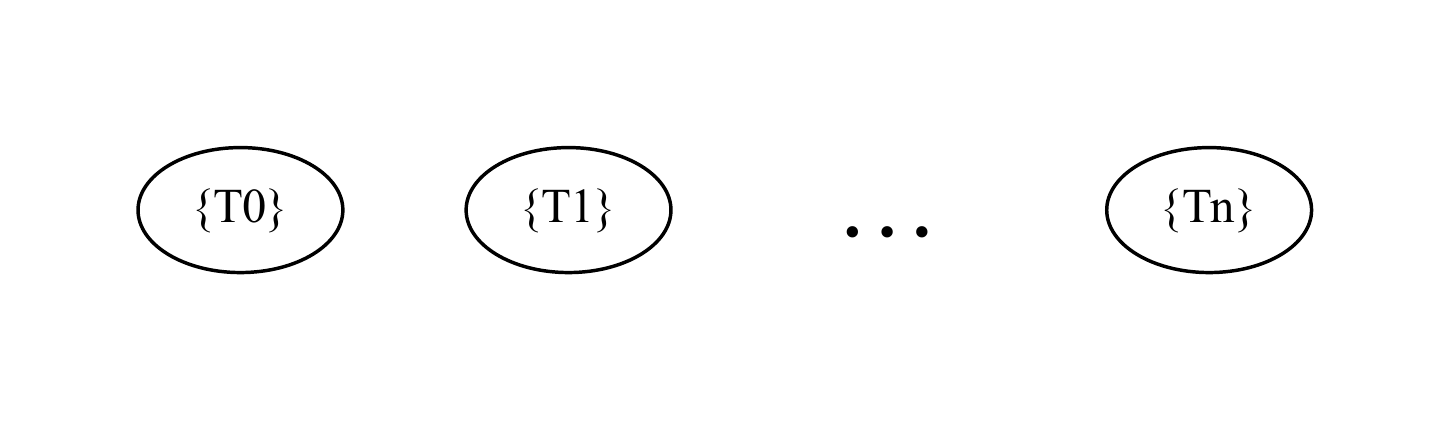}}}
         \caption{$\doadep = n$}
         \label{sfig:graph-indep}
     \end{subfigure}
        \caption{Abstract dependency graphs (DGs) and their respective task
        dependency degree of asynchrony, $\doadep$. Abstract because only task
        sets (nodes), their orderings (indices) and dependencies (edges) are
        given. Those DGs contain no information about the number of tasks in
        each task set, their resource requirements, task execution time (TX),
        \etc~Task set indices are ordered in a breadth-first manner.
        }
        \label{fig:dags}
\end{figure}

Asynchronous task execution only applies to workflows with a dependency graph
(DG) that contains one or more forks, with diverging paths. As such, given
enough resources, independent tasks of different types can execute
asynchronously. Figure~\ref{fig:dags} shows four DGs representative of four
workflows with varying levels of inter-task data dependence. Note that task set
indices are ordered breadth-first, and that the y-axis of our DG does not
correlate with time.

We define the degree of asynchronicity ($\doadep$) as the number of independent
execution branches minus 1. To discover independent branches, we perform a depth
first search on the DG. The lower bound on  ($\doadep$), permitted by task
dependencies, arises when the DG is sequential, \eg, a linear chain,
Fig.~\ref{sfig:graph-chain} for which $\doadep = 0$. The upper bound, \ie,
completely asynchronous, arises a DG whose edge set is the empty set such as
that in Fig.~\ref{sfig:graph-indep}, where $\doadep = n$. In between those
bounds, asynchronicity varies as a function of the number of forks with
diverging paths.

\subsection{Condition II: Resource Availability}\label{ssec:cond-resreqs}

Assuming unlimited resources, inter-task dependency is, in principle, the
condition for determining the degree of asynchronicity for a given workflow. For
example, the DG in Fig.~\ref{sfig:graph-indep} has $\doadep = n$, so in the
case of infinite resources, each of the $n+1$ task sets' tasks can execute
asynchronously. In practice, however, resources are limited, and therefore an
additional condition based on the available (allocated) resources must be
imposed.

To describe the resource-permitted degree of asynchronicity, $\doares$, consider
again the
DG in Fig.~\ref{sfig:graph-indep}. Let $\widetilde{R}$ be the set of all
allocated resources and $\cup_i R_i$ be the set resources required to
asynchronously execute the $n+1$ task sets. Complete asynchronicity is achieved
if and only if $\cup_i R_i \subseteq \widetilde{R}$, in which case $\doares >=
\doadep = n$. On the other extreme, if $R_i = \widetilde{R}~\forall_i$, then we
have two possibilities: (1) execute each task set in arbitrarily or based on a
given scheduler's heuristics, since the execution of the full set of tasks from
any task set requires 100\% of the allocated resources; or (2) execute from two
or more task sets the subsets such that $\cap_i R_i = \widetilde{R}$. Despite
$\doares = 0$ in the first case and $\doares > 0$ in the second, there exists an
equivalence between the two wherein the latter effectively is the collapse of a
dependency-free DG to a chain as in Fig.~\ref{sfig:graph-chain}. This scenario
highlights how haphazard attempts to adopt asynchronicity into a workflow will
not always yield any benefit for the workflow as a whole, and instead can lead
to significant loss of development time during the design process.

Richer, albeit more complex, situations lie between the two extremes. Take, for
example, the workflow DG of Fig.~\ref{sfig:graph-doa4}, which has $\doadep =
4$. Depending on the resource requirements for each of $T_2-T_9$ task sets, $0
\leq \doares \leq 5$. As such, we define,
\begin{equation}
    \text{WLA} \equiv \min[\doadep, \doares]
\end{equation}
to quantify the asynchronicity permitted by the workflow.

Note that asynchronous executions are not necessarily concurrent, and
synchronous executions are not necessarily sequential. Tasks are asynchronous
when they can execute independently of one another, while tasks are concurrent
when they execute at the same time. Thus, asynchronous executions can be either
sequential or concurrent, depending on resource availability. For example, two
independent tasks (${T_O}$ and ${T_1}$ from Fig.~\ref{sfig:graph-doa4}) can be
run one after the other if not enough resources are available to execute them
concurrently. Analogously, two tasks that depend on each other (\eg inter-task
communication dependency) can execute concurrently when enough resources are
available.

\subsection{Benefits of Workflow-level Asynchronicity}\label{ssec:benefits}

A primary motivation for asynchronous task execution is to improve resource
utilization which, in turn, correlates to task throughput and thus workflow
makespan. However, examples have already been given that show how the
introduction of asynchronicity does not guarantee increased task throughput,
specifically when the majority of upstream tasks (ancestors) dominate resource
use. When a workflow requires resources with a $\doares > 0$ and it is possible
to exploit TX \textit{masking}, the overall workflow makespan can be reduced by
achieving higher task throughput.


TX masking happens when longer running tasks effectively `hide' contributions
from shorter running tasks to a workflow's makespan (TTX). To illustrate
masking, consider the workflow DG of Fig.~\ref{sfig:graph-doa1} for which
$\doadep = 1$. Assume the allocated resources are such that once $T_0$
completes, the chains $\{T_1, T_3, T_5\}$ and $\{T_2, T_4\}$ can execute
asynchronously; then $\doares = \doadep = 1$. For concreteness, further assume
the following TTX assignments:
\begin{equation*}
    t_0 = 500\text{s},~t_1 = t_2 = 1000\text{s},~2t_3 = 2t_5 = t_4 = 4000\text{s}.
\end{equation*}

To demonstrate the utility of masking, we begin with a purely sequential
workflow using the Pipeline, Stage and Task (PST)
model~\cite{balasubramanian2018harnessing}, where the DG represents a pipeline,
each rank corresponds to a stage, and each task set to tasks. In the sequential
model, the TTX is given by,
\begin{equation}
    \tseq = \sum_i t_i + C, \label{eqn:tseq}
\end{equation}
where $i$ represents $i$th stage and $t$ represents TX of given stage, and $C$
is a constant representing the workflow management system's overheads such as
communications. In general, $C$ is negligible with task execution times $O(10)$
minutes and larger. As such, we disregard small overheads in future calculations
but consider them in later analyses. Plugging into Eqn.~\ref{eqn:tseq} the above
assigned TX values, the sequential TTX is $t^{\text{Seq}} = 7500\text{s}$.

We now demonstrate how the workflow TTX can be significantly reduced by
exploiting the conditions for asynchronicity and the parameters of the workflow;
specifically here, $\text{WLA} = \min(\doadep, \doares) = 1$ and the TX
assignments are the same as above. First, the WLA enables asynchronous execution
of the chains $H_1 \equiv \{T_1, T_3, T_5\}$ and $H_2 \equiv \{T_2, T_4\}$.
Second, the TX assignments lead to masking such that the contribution from task
set $T_5$ to the workflow TTX is hidden by the execution of task set $T_4$ since
$t_4 = t_3 + t_5$ and the corresponding task sets now execute asynchronously. In
general, the asynchronous TTX for an asynchronous workflow is given by,
\begin{align}
    \label{eqn:tasync}
    \tasync &= \sum_{i} t_i + \max[tt_{H_j}] + C \\
            &\leq \tseq, \nonumber
\end{align}
where $t_i$ represents the sequential task sets and $tt_{H_j}$ represents the
TTX of each independent branch.
\begin{align}
    \label{eqn:tth}
    tt_{H_j}&= \sum_{j} t_j  \\
            &\leq \tseq, \nonumber
\end{align}

We have equality if and only if $t_{H_j} = t_{H_k}~\forall_{j,k}$. Asynchronous
execution of the workflow under discussion therefore has a TTX of $t_0 + t_{H_1}
= 5500\text{s}$. We define the relative improvement as,
\begin{align}
    I &\equiv 1 - \frac{\tasync}{\tseq},
    \label{eqn:relimp}
\end{align}
for which, in this example, gives roughly a 26\% reduction compared to the
sequential execution. This example makes clear the potential gains when a
workflow is designed to leverage WLA as well as TX masking.

%% file: experiments.tex

We design three experiments to characterize the performance of asynchronous task
execution, using three heterogeneous workflows with different inter-task
dependencies. We execute two implementation of a workflow for each experiment,
one asynchronous and one sequential, and compare their performance. All
workflows' task executes the \texttt{stress} application and, for simplicity, we
omit operations such as data movement and staging which are largely
platform-dependent. \texttt{Stress} allows implementing task heterogeneity by
setting arbitrary task duration (TX) and resource requirements (number of
CPUs/GPUs). Further, using a synthetic task executable allows to make judicious
use of shared allocations and alleviate overheads, \eg, having to use
application-specific input files, configurations, \etc. As we focus on workflow
asynchronicity, this approach permits analysis without loss of generality.


\subsection{DeepDriveMD}\label{ssec:ddmd}

DeepDriveMD~\cite{brace2021achieving} comprises four types of tasks---\simtask,
\aggtask, \mltask, and \inftask---as described earlier. All tasks but \aggtask{}
are heterogeneous, \ie, both CPUs and GPUs, and the number of each type of task
and their execution times vary among task sets. Table~\ref{tab:wf-ddmd} details
the workflow parameters which are scaled appropriately with respect to the
number of nodes used for these studies.

\begin{table}
\caption{DeepDriveMD workflow tasks. The TX for each task was extracted
from~\cite{brace2021achieving} and scaled down by a factor of four. A variable
offset $0.05\sigma$ is added to TX of each task to mimic the stochastic behavior
of actual executables.}
\label{tab:wf-ddmd}
\centering
\begin{tabular}{l|r|r|r|r}
\toprule
Task Set                   &
CPU cores/Tasks            &
GPUs/Tasks                 &
\# Tasks ($\times 3$)      &
TX ($\pm 0.05\sigma$) [s] \\
\midrule
Simulation                 &
4                          &
1                          &
96                         &
340                        \\
Aggregation                &
32                         &
0                          &
16                         &
85                         \\
Training                   &
4                          &
1                          &
1                          &
63                         \\
Inference                  &
16                         &
1                          &
96                         &
38                         \\
\bottomrule
\end{tabular}
\end{table}

There exist dependencies between the different task sets (similar to
Fig.~\ref{sfig:graph-chain}); \simtask{} tasks are responsible for producing the
data that is consumed by \aggtask{} tasks, the aggregated data is then inputted
to ML model \mltask{} tasks, and lastly \inftask{} tasks perform inference using
the trained models. As such, a single iteration of this workflow necessarily
needs to be sequential in terms of the execution of various tasks. In practice,
however, ML training and inference typically require multiple iterations
(sufficient data volumes) for convergence and to reach some level of accuracy or
precision in terms of their predictions. It is therefore possible and, as will
be demonstrated, advantageous to asynchronously execute tasks of different types
among iterations while also ensuring dependencies are met.

Each task set can execute concurrently, \textit{e.g.}, all \simtask{} tasks run
at the same time, each task set is executed sequentially, as described
previously. As such, sequential execution of this workflow uses a single
pipeline to orchestrate task execution, where each stage executes sequentially
and spawns at least one task that receives input from a preceding one. To
construct an asynchronous workflow from multiple (three) independent chains, we
use the equivalent DG shown in Fig.~\ref{sfig:graph-ddmd}\footnote{An
alternative DG representation of this workflow is three independent chains, one
for each iteration. However, this would not admit asynchronous execution as
defined in \S\ref{sec:design}.}, where task sets are staggered. Using the PST
nomenclature, the complete DG corresponds to a single pipeline, each rank to a
stage, and each task set to the tasks comprising the workload. Task
specifications for this workflow are given in Table~\ref{tab:wf-ddmd}.

As a future work, we will remove dependencies between independent chains
caused by putting each rank in a stage. In that context, we will focus on
task-level dependencies to achieve adaptive asynchronicity, where: (1) tasks
from different, non-converging branches can execute completely asynchronously
(\eg~from fig.~\ref{sfig:graph-ddmd} $Arrg_0$ and $Train_1$ can run at the same
time); and (2) tasks from different converging branches can still execute
asynchronously as long as they don't have any dependencies between each other
(\eg~from fig.~\ref{sfig:graph-doa2} $T_1$ and $T_5$ can execute asynchronously
and possibly concurrently).

\subsection{Abstract-DG}\label{ssec:abstract-dag}

\begin{figure}[!ht]
     \centering
     \begin{subfigure}[b]{0.5\columnwidth}
         \centering
         \includegraphics[width=\columnwidth]{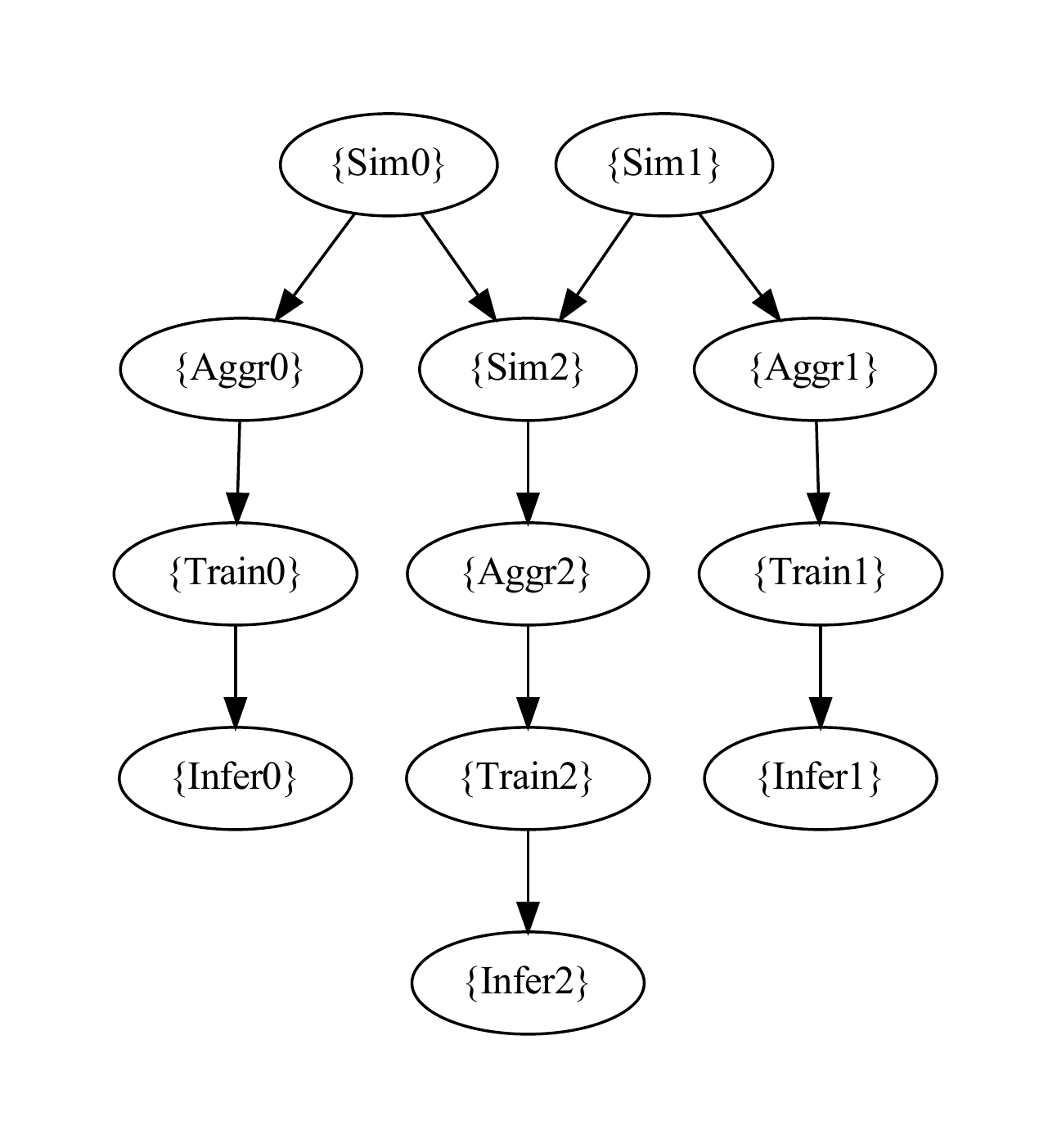}
         \caption{}\label{sfig:graph-ddmd}
     \end{subfigure}%
     \hfill
     \begin{subfigure}[b]{0.5\columnwidth}
         \centering
         {\raisebox{10pt}{\includegraphics[width=\columnwidth]{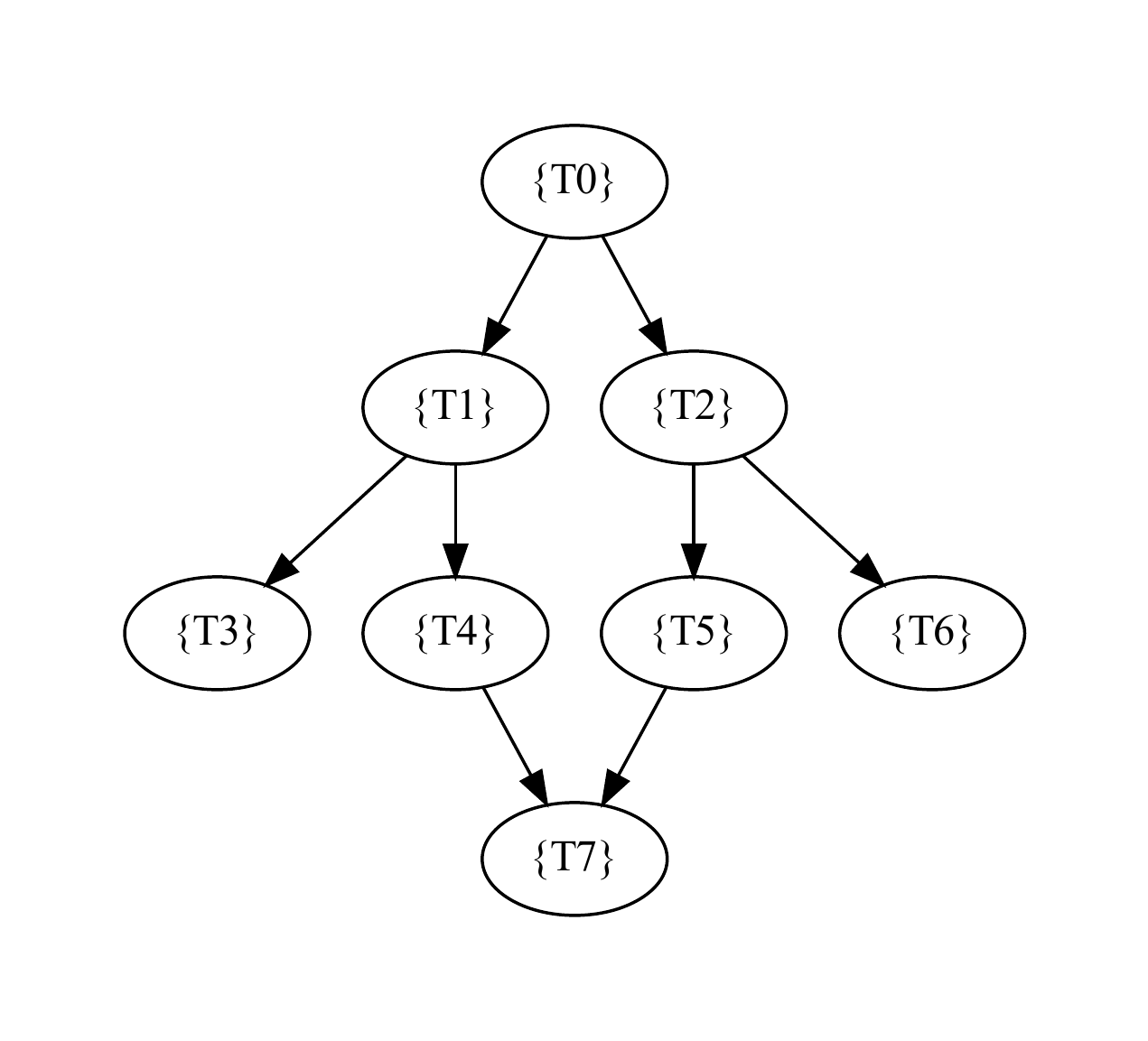}}}
         \caption{}\label{sfig:graph-doa2}
     \end{subfigure}
        \caption{Dependency graphs corresponding to (a) DeepDriveMD and (b)
        arbitrary workflows.}\label{fig:dags2}
\end{figure}

We constructed two additional workflows from a single arbitrary abstract DG.
Figure~\ref{sfig:graph-doa2} shows the DG considered for these studies, which
consists of eight task sets labeled T0--T7 and dependencies among them.

Two concrete workflows based on the abstract DG, denoted c-DG1 and c-DG2, are
considered to support the conditions for WLA, and to quantify the degree to
which asynchronicity is permitted by a specific workflow instantiation. Both
concrete workflows are assigned task sets with an arbitrary number of tasks,
resources and TX; see Table~\ref{tab:wf-adag}.

\begin{table}
    \caption{Summary of the abstract DG workflow tasks for the concrete DGs `c-DG1'
    and `c-DG2'. Similar task types are assigned the same resources and their
    respective task sets are grouped within braces. The task set TX values are
    sampled from a normal distribution $\mathcal{N}(\mu, \sigma=0.05)$, where $\mu =
    (\text{Mean TTX Fraction} \times 2000~\text{[s]})$ to emulate stochastic
    behavior of actual executables; these fractions have been rounded to two
    significant figures in the table.}\label{tab:wf-adag}
    \centering
    \begin{tabular}{l|r|rr|rr|rr}
    \toprule
    \multirow{2}*{Task Set}                    &
    \multirow{2}*{CPU cores/Task}              &
    \multicolumn{2}{c|}{GPUs/Task}             &
    \multicolumn{2}{c|}{\# Tasks ($\times 3$)} &
    \multicolumn{2}{c}{Mean TTX Fraction} \\
                                            &
                                            &
    c-DG1                                      &
    c-DG2                                      &
    c-DG1                                      &
    c-DG2                                      &
    c-DG1                                      &
    c-DG2                                      \\
    \midrule
    T0                     &
    16                     &
    1                      &
    1                      &
    96                     &
    96                     &
    0.38                   &
    0.19                   \\
    \{T1, T2\}             &
    40                     &
    0                      &
    0                      &
    32                     &
    32                     &
    0.11                   &
    0.08                   \\
    \{T3, T6\}             &
    4                      &
    0                      &
    1                      &
    16                     &
    96                     &
    0.06                   &
    0.38                   \\
    \{T4, T5\}             &
    32                     &
    1                      &
    1                      &
    16                     &
    16                     &
    0.08                   &
    0.12                   \\
    T7                     &
    4                      &
    1                      &
    0                      &
    96                     &
    16                     &
    0.36                   &
    0.23                   \\
    \bottomrule
    \end{tabular}
\end{table}

Task properties are chosen for each concrete workflow to explore two distinct
use-cases from a single abstract DG, and how different parameters can affect the
relative improvement when incorporating asynchronicity. Note that, contrary to
DeepDriveMD's DG, each rank is not associated with a stage. As mentioned above,
in an adaptive asynchronous execution, tasks (T1, T4) and (T2, T5) would execute
asynchronously since they do not have any dependencies, but T7 would only
execute after both T4 and T5 are finished.

%% file: performance.tex

We evaluate the achievable level of asynchronicity, resource utilization, and
TTX for the experimental workflows described in \S\ref{sec:experiments},
executing them on the Oak Ridge National Laboratory's Summit supercomputer.
Summit has roughly 4600
compute nodes, each with 2 24 cores Power9 CPU and six NVIDIA V100 GPUs.
In these studies, we use a total of 16 nodes for each experiment, giving 706
usable CPU core (62 cores are reserved by the system) and 96 GPUs.

We construct synthetic workloads representative of real-world workflow
applications to characterize and evaluate the performance of our experiments.
For example, we consider workloads comprising a combination of homogeneous and
heterogeneous tasks, and different types of tasks with varying TX.
To calculate the relative improvement $I$ (Eqn.~\ref{eqn:relimp}), for each
experiment, we execute both sequential and asynchronous modes. We impose a TTX
constraint on c-DG1 and c-DG2; the sequential TTX is about $2000~\text{s}$ for
both.
Results for asynchronous executions are summarized at the end of this section,
in Table~\ref{tab:results}.
    As part of our future work, we will test real-life workflows from different scientific domains on various HPC facilities.  


\begin{table*}[h]
    \caption{Summary of experimental results. Predicted values of $\tseq$,
    $\tasync$, and $I$ include corrections accounting for overheads in EnTK ($4\%$)
    and those incurred from enabling asynchronicity ($2\%$) into the framework.
    While the degrees of asynchronicity ($\doadep$ and $\doares$)
    directly determine the workload-level asynchronicity (WLA) for a given workflow,
    resource utilization and task throughput, captured by the relative improvement
    ($I$), depend on available resources and TX masking.}\label{tab:results}
    \centering
    \begin{tabular}{l|rrr|rr|rr|rr}
    \toprule
    \multirow{2}*{Experiment}                   &
    \multirow{2}*{$\doadep$}                    &
    \multirow{2}*{$\doares$}                    &
    \multirow{2}*{$\text{WLA}$}                 &
    \multicolumn{2}{c|}{$\tseq$ [s]}             &
    \multicolumn{2}{c|}{$\tasync$ [s]}           &
    \multicolumn{2}{c}{$I$} \\
                                                &
                                                &
                                                &
                                                &
    %
    Pred.                                   &
    Calc.                                    &
    Pred.                                   &
    Calc.                                    &
    Pred.                                   &
    Calc.                                     \\

    \midrule
    DeepDriveMD   &
    2             &
    1             &
    1             &
    1578          &
    1707          &
    1399          &
    1373          &
    $0.113$       &
    $0.196$       \\
    c-DG1         &
    2             &
    2             &
    2             &
    2000          &
    $1945$        &
    1972          &
    $1975$        &
    0.014         &
    $-0.015$      \\
    c-DG2         &
    2             &
    2             &
    2             &
    2000          &
    1856          &
    1378          &
    1372          &
    0.311         &
    $0.261$       \\
    \bottomrule
    \end{tabular}
\UP
\end{table*}

\subsection{DeepDriveMD}\label{ssec:perf-ddmd}

\begin{figure*}[!ht]
     \centering
     \begin{subfigure}[b]{0.5\textwidth}
         \centering
         \includegraphics[width=\textwidth]{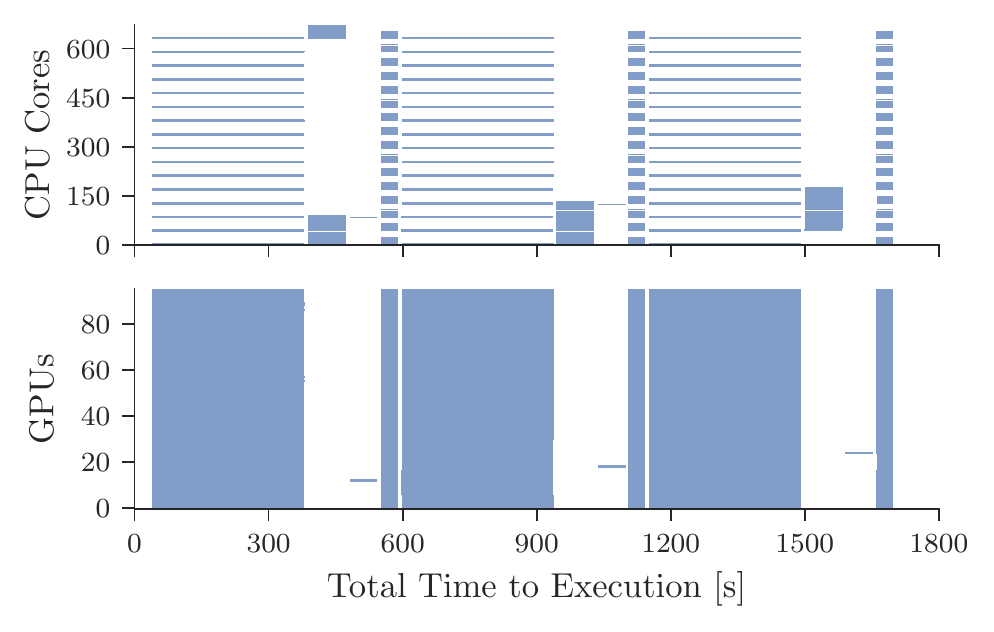}
         \caption{Sequential ($1707~\text{s}$)}\label{sfig:ddmd-seq-resuse}
     \end{subfigure}%
     \begin{subfigure}[b]{0.5\textwidth}
         \centering
         \includegraphics[width=\textwidth]{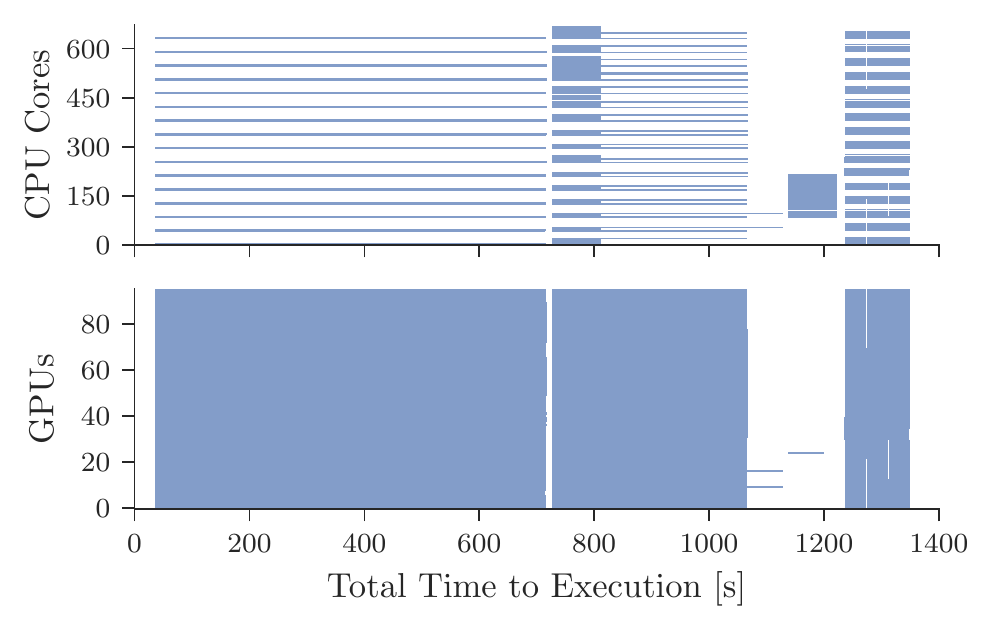}
         \caption{Asynchronous ($1373~\text{s}$)}\label{sfig:ddmd-async-resuse}
     \end{subfigure}
        \caption{Three iterations of DeepDriveMD workflow, with sequential
        (left) and asychronous (right) execution. The TTX improvement achieved
        with asynchronicity is about $20\%$.}\label{fig:resuse-ddmd}
\end{figure*}

In principle, the DeepDriveMD
workflow is sequential: $\simtask \rightarrow
\aggtask \rightarrow \mltask \rightarrow \inftask$. 
To implement asynchronicity, we started multiple executions of the DeepDriveMD workflow with different starting times so to run different stages asynchronously. 
As such, the workflow TTX is
bounded below by the level of asynchronicity permitted by the number of times it
is executed. Inspection of the DG in Fig.~\ref{sfig:graph-ddmd} reveals the
workflow (each task set) is executed three times, and three independent chains
beginning at rank 1 means $\doadep = 2$. Based on task requirements in
Table~\ref{tab:wf-ddmd}, the allocated resources give $\doares = 1$. Hence,
$\text{WLA} = 1$ for this workflow.

Figure~\ref{fig:resuse-ddmd} shows the CPU and GPU resource utilization for
three iterations of the DeepDriveMD workflow as a function of TTX, with
white-space representing resources on which there are no executing tasks and
colored regions representing utilized resources. Comparing
Fig.~\ref{sfig:ddmd-seq-resuse} and Fig.~\ref{sfig:ddmd-async-resuse},
sequential and asynchronous execution, respectively, it is clear the allocated
resources are far better utilized with asynchronicity. The improved use of
resources naturally leads to a higher task throughput, as shown by the decrease
in the asynchronous workflow TTX. From Eqn.~\ref{eqn:tseq}, with the TX values
from Table~\ref{tab:wf-ddmd}, sequential execution has a TTX of approximately
$3\tseq = 1578~\text{s}$, the leading factor a result of the workflow being
executed three times. The independent chain in the asynchronous workflow DG with
the largest combined TX is the chain with task set $\{\text{Sim}2\}$ as the
top-level ancestor, giving $t_{H} = \tseq = 526~\text{s}$. Task sets
$\{\text{Sim1}\}$, $\{\text{Sim2}\}$, and the three \inftask{} task sets, each
requiring all 96 GPUs, are ineligible for asynchronicity due to insufficient
resources. Then, plugging into Eqn.~\ref{eqn:tasync} the values $2t_\text{Sim0}
= 2t_\text{Sim1} = 680~\text{s}$, $3t_\text{Infer0} = 3t_\text{Infer1} =
3t_\text{Infer2} = 114~\text{s}$ and $t_{H}$, the asynchronous workflow
execution time is calculated to be $\tasync = 1320~\text{s}$, yielding $I=0.17$.

Important to note is the discrepancy of about $4\%$ between the calculated
$\tasync$ and measured value of $1373~\text{s}$. This can be attributed to two
key facts: (1) corresponding to roughly half the disagreement is we disregard
constant overheads in the EnTK framework (the constant factor $C$ in
Eqn.~\ref{eqn:tseq}), and (2) Eqn.~\ref{eqn:tasync} assumes infinite resources.
Since each \inftask{} task set requires all available resources, we have the
scenario described in Sec.~\ref{ssec:cond-resreqs} where the otherwise
independent chains collapse to a single chain dependent on the resource
allocation.

In light of the TTX disagreement between theory and practice for this
multi-iteration workflow, an alternative formulation for calculating $\tasync$
can be used:
\begin{align}
    \tasync &= n \tseq - (n-1) \taggr - (n-2) \tml.
    \label{eqn:tasync-mask}
\end{align}
such that $n-1$ ($n-2$) sets of \aggtask{} (\mltask) tasks can be executed
concurrently across iterations. Equation~\ref{eqn:tasync-mask} makes clear that
TX-masked tasks do not contribute to the overall TTX (they are subtracted away).
Thus, unlike Eqn.~\ref{eqn:tasync}, Eqn.~\ref{eqn:tasync-mask} correctly does
not mask $\tinf$ due to resource constraints. It will be useful in later
analysis to generalize this equation as,
\begin{align}
    \tasync &= \sum_{k=0}^{-1+n} t_k (-k + n).
    \label{eqn:tasync-mask-gen}
\end{align}
In the DeepDriveMD workflow, \simtask{} tasks have sufficiently large TX such
that \aggtask{} and \mltask{} tasks can both be masked, \ie, $t_\text{Sim2} >
t_\text{Aggr0} + t_\text{Train0} = t_\text{Aggr1} + t_\text{Train1}$.
Substituting the respective values into Eqn.~\ref{eqn:tasync-mask}, the
estimated TTX is $1345~\text{s}$, a difference of $2\%$ with respect to the
measured TTX the remaining. This remaining time is attributed to the constant
overheads from the framework, visible in Fig.~\ref{fig:resuse-ddmd} as
white-space between task set executions.

\begin{figure*}[!h]
     \centering
     \begin{subfigure}[b]{0.5\textwidth}
         \centering
         \includegraphics[width=\textwidth]{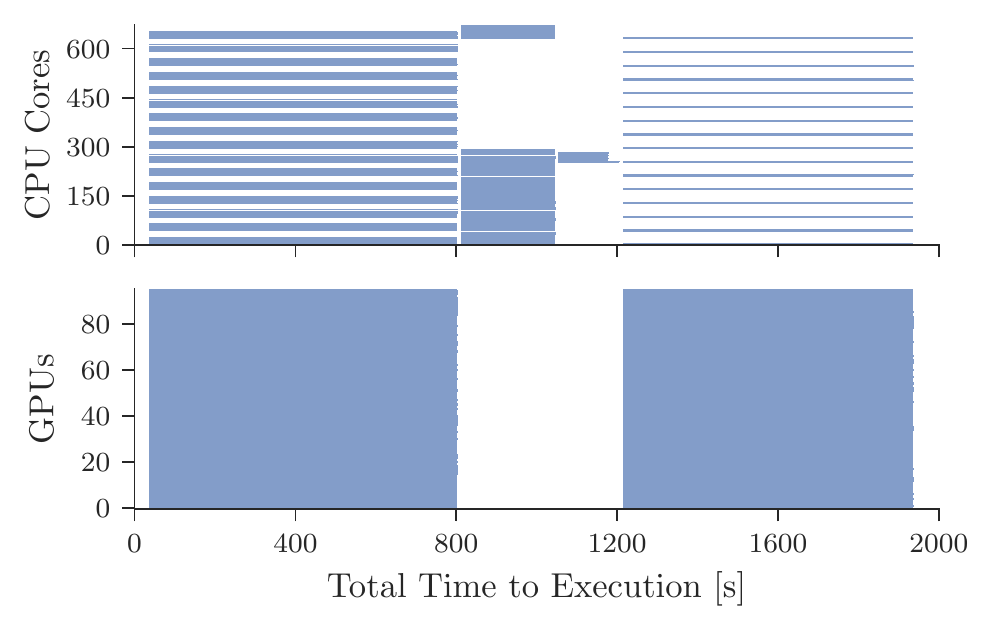}
         \caption{Sequential ($1945~\text{s}$)}\label{sfig:ddmd-seq-resuse}
     \end{subfigure}%
     \begin{subfigure}[b]{0.5\textwidth}
         \centering
         \includegraphics[width=\textwidth]{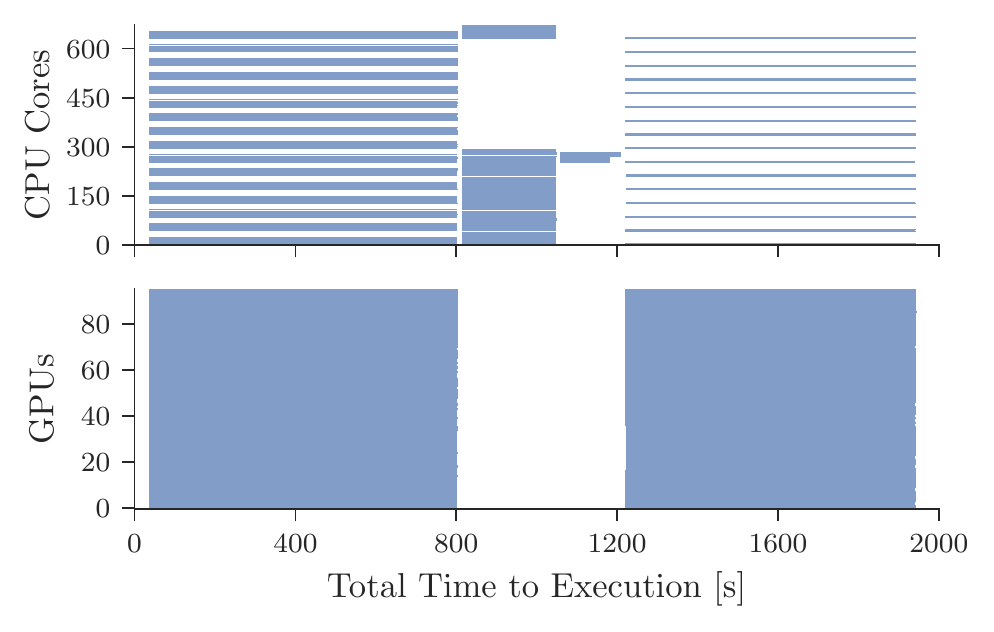}
         \caption{Asynchronous ($1975~\text{s}$)}\label{sfig:ddmd-async-resuse}
     \end{subfigure}
    \caption{Resource utilization for workflow c-DG1. Sequential (left) and
    asynchronous (right) execution. Due to the small TX (6\% and 7\%) of the
    task sets permitting asynchronous execution ($\{T_3,T_6\}$ and $\{\{T_4,
    T_5\}, T_7\}\}$), and additional overheads in spawning multiple workflow
    executions required for asynchronicity, resource utilization and task
    throughput improvements are negligible, ultimately resulting in a negative
    value of $I$.}\label{fig:resuse-adag1}
\end{figure*}

\subsection{c-DG1}\label{ssec:perf-adag1}

The task property assignments in c-DG1 were chosen to demonstrate a case when
asynchronicity has a negative impact on a workflow's TTX and task throughput.
The workflow is represented by the DG in Fig.~\ref{sfig:graph-doa2}, which has
$\doadep = \doares = 2$ for $\text{WLA} = 2$. As such, the workflow permits
asynchronicity, however, the task requirements, given in
Table~\ref{tab:wf-adag}, prevent an improvement in resource utilization and
therefore task throughput.

This is due to the fact that, while it is possible to run together task sets
$\{T_3,T_6\}$ and $\{\{T_4, T_5\}, T_7\}\}$, the relatively small TX of
$\{T_3,T_6\}$ and $\{T_4, T_5\}$, as well as dependencies for $\{T_7\}$ to begin
executing, offer a negligible improvement with TX masking. Indeed, the
additional overheads---roughly 2\% of the workflow TTX---induced by introducing
asynchronicity here result in a larger TTX than that of the sequential
execution. Direct application of Eqn.~\ref{eqn:tasync} yields a predicted
asynchronous TTX of $1860~\text{s}$, which disagrees with the measured TTX by
just under $6\%$: $4\%$ of the difference accounted for by earlier arguments
from the DeepDriveMD example, and a nearly $2\%$ performance hit for overheads
incorporating asynchronicity. Accounting for this difference,
Eqn.~\ref{eqn:relimp} can be used to estimate the potential gains, giving $I =
0.01$, which does not provide motivation to adopt asynchronicity. Using the
measured values of the sequential and asynchronous versions of this workflow,
$I=-0.015$; asynchronicity indeed has a negative relative improvement.
Therefore, workflows with similar traits to c-DG1 are preferentially sequential.

\begin{figure*}[!h]
     \centering
     \begin{subfigure}[b]{0.5\textwidth}
         \centering
         \includegraphics[width=\textwidth]{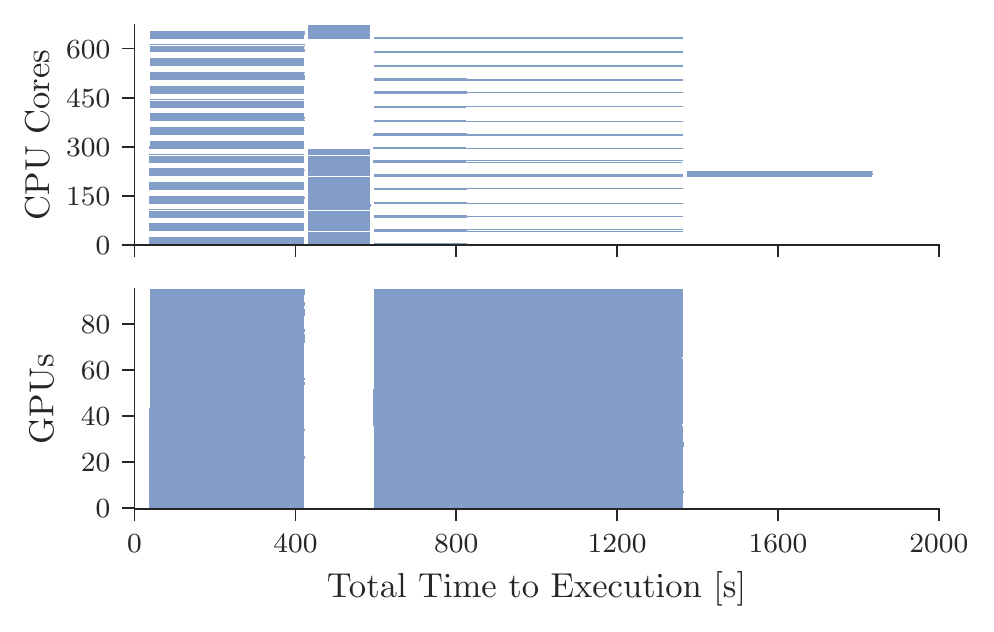}
         \caption{Sequential ($1856~\text{s}$)}\label{sfig:ddmd-seq-resuse}
     \end{subfigure}%
     \begin{subfigure}[b]{0.5\textwidth}
         \centering
         \includegraphics[width=\textwidth]{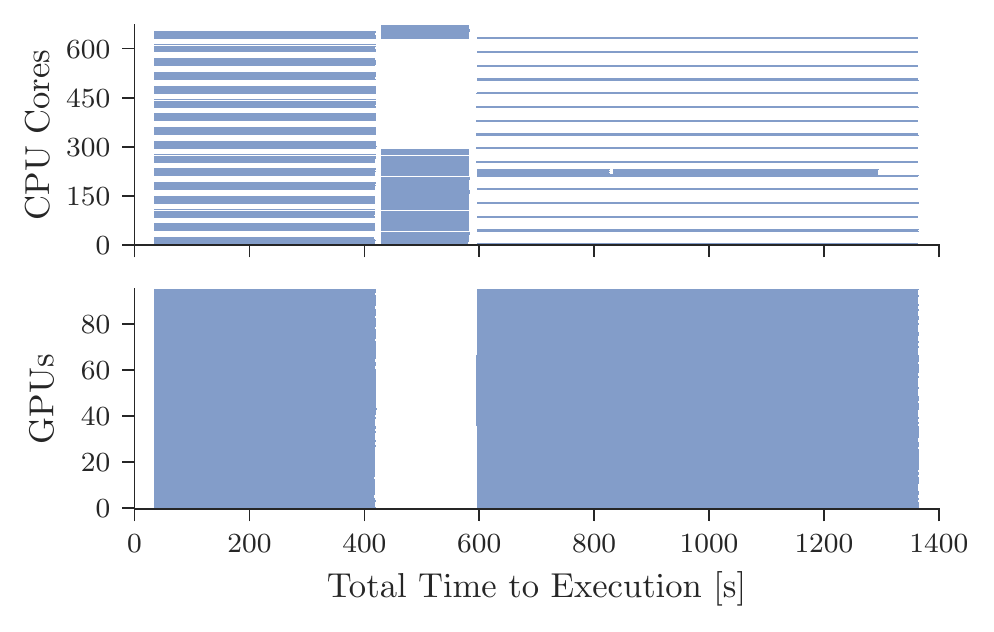}
         \caption{Asynchronous ($1372~\text{s}$)}\label{sfig:ddmd-async-resuse}
     \end{subfigure}
    \caption{Resource utilization for workflow c-DG2. Sequential (left) and
    asynchronous (right) execution, with independent task sets' TTX such that
    $t_{\text{T3,T6}} \sim t_{\text{T4,T5}} + t_{\text{T7}}$. Exploiting TX
    masking and the fact that sufficient resources are available, the workflow
    TTX is reduced by nearly $26\%$.}
    \label{fig:resuse-adag2}
\end{figure*}

\subsection{c-DG2}\label{ssec:perf-adag2}

We demonstrate last a concrete workflow from the abstract DG in
Fig.~\ref{sfig:graph-doa2}. Let us take the proposed approach of first
predicting the relative improvement for this workflow using the relevant task
parameters from Table~\ref{tab:wf-adag} for c-DG2. From earlier analyses, we can
anticipate an under-estimate of $4-6\%$ using Eqn~\ref{eqn:tasync}. Plugging in
the task execution times, the predicted TTX of this workflow is calculated to be
$1300\text{s}$. Assuming an under-estimate, we account for a $6\%$ discrepancy
with the measured TTX to give $1378~\text{s}$. From Eqn.~\ref{eqn:tseq}, the
predicted TTX of the sequential execution of c-DG2 is $2000~\text{s}$. The
estimated relative improvement is thus $I = 0.31$, suggesting the workflow will
benefit from asynchronicity. Experimentation measures sequential and
asynchronous execution of $1856~\text{s}$ and $1372~\text{s}$, respectively, for
$I = 0.26$. The non-zero $\doadep$ and $\doares$, which imply a WLA $>0$, as
well as the agreement between the predicted and measured asynchronous TTX
guarantee an asynchronicity advantage.

%% file: conclusions.tex

We studied asynchronous execution 
and analyzed the
performance impact that diverse degrees of asynchronicity have when executing
workflows with heterogeneous tasks on HPC platforms, focusing on ML-coupled
workflows. This paper offers four contributions: (1) an asynchronous
implementation of DeepDriveMD, a framework to execute ML-driven
molecular-dynamics workflows on HPC platforms at scale; (2) a performance
evaluation of asynchronous DeepDriveMD; (3) a model of asynchronous behavior;
and (4) a general performance evaluation of that model for workflows with
varying degrees of asynchronous execution.

By employing asynchronous execution in DeepDriveMD, a relative improvement
$I=0.196$ over the sequential mode was achieved. This improvement translates to
a reduction of an experimentally measured sequential TTX of $1707~\text{s}$ to
$1372~\text{s}$ with asynchronicity. Two concrete workflow realizations were
defined from a single arbitrary abstract DG to demonstrate a case where the
introduction of asynchronicity has a negative effect on TTX, $I=-0.015$, and
another that gives considerable asynchronous advantage, $I=0.261$. In all
experiments, our model presented Sec.~\ref{sec:conds} accurately predicted
within less than $6\%$ the experimentally measured values modulo framework
constant overheads.




The contributions of this paper offer a first step towards the development of
workflow management capabilities to execute heterogeneous tasks at scale on HPC
platforms, with higher resource utilization and lower makespan. Those
capabilities should be tailored to the specific requirements of ML-driven
(heterogeneous) HPC workflows, and offer an
asynchronous execution capability.
That is a necessary precondition for the efficient and effective deployment of
ML-driven workflows on HPC platforms at scale.


In the future, we are looking to improve asynchronicity using adaptive
asynchronous execution of ML-driven HPC workflows. Adaptive execution will allow
varying the degree of asynchronous execution at runtime, based on dynamic
requirements.


